\documentclass[
    ,final            % use final for the camera ready runs
  ]
  {aipproc}

\usepackage{axodraw,amsmath}

\layoutstyle{8x11double}

\graphicspath{{figs/}}

\newcommand{\FUTA}{{\bf FUTA}}
\newcommand{\FUTB}{{\bf FUTB}}

\def\reffi#1{\mbox{Fig.~\ref{#1}}}

\def\citere#1{\mbox{Ref.~\cite{#1}}}
\def\citeres#1{\mbox{Refs.~\cite{#1}}}

\def\order#1{${\cal O}(#1)$}

\newcommand{\MZ}{M_Z}
\newcommand{\MA}{M_A}

\newcommand{\Mh}{M_h}

\newcommand{\mt}{m_{t}}

\newcommand{\mb}{m_{b}}
\newcommand{\mbbar}{\overline{m}_b}

\newcommand{\db}{\De_b}

\newcommand{\tsf}{\theta\kern-.20em_{\tilde{f}}}
\newcommand{\tsfp}{\theta\kern-.20em_{\tilde{f}\prime}}
\newcommand{\tsq}{\theta\kern-.15em_{\tilde{q}}}

\newcommand{\VL}{\left( \begin{array}{c}}
\newcommand{\VR}{\end{array} \right)}
\newcommand{\ML}{\left( \begin{array}{cc}}
\newcommand{\MLd}{\left( \begin{array}{ccc}}
\newcommand{\MLv}{\left( \begin{array}{cccc}}
\newcommand{\MR}{\end{array} \right)}

\newcommand{\tb}{\tan \beta}

\newcommand{\tev}{\,\, \mathrm{TeV}}
\newcommand{\gev}{\,\, \mathrm{GeV}}

\newcommand{\BC}{\begin{center}}
\newcommand{\EC}{\end{center}}
\newcommand{\BE}{\begin{equation}}
\newcommand{\EE}{\end{equation}}
\newcommand{\BEA}{\begin{eqnarray}}
\newcommand{\BEAnn}{\begin{eqnarray*}}
\newcommand{\EEA}{\end{eqnarray}}
\newcommand{\EEAnn}{\end{eqnarray*}}

\newcommand{\id}{{\rm 1\kern-.12em
\rule{0.3pt}{1.5ex}\raisebox{0.0ex}{\rule{0.1em}{0.3pt}}}}
\newcommand{\lsim}
{\;\raisebox{-.3em}{$\stackrel{\displaystyle <}{\sim}$}\;}

\def\ga{\gamma}

\def\De{\Delta}

\newcommand{\br}{{\rm BR}}

\newcommand{\amu}{a_\mu}

\begin{document}

\onecolumn

\thispagestyle{empty}
\setcounter{page}{0}
\def\thefootnote{\fnsymbol{footnote}}

\begin{flushright}
\mbox{}
arXiv:0809.2397 [hep-ph]
\end{flushright}

\vspace{1cm}

\begin{center}

{\fontsize{15}{1} 
\sc {\bf Selecting Finite Unified Theories with Current Data}}
\footnote{talk given at the {\em SUSY\,08}, 
June 2008, Seoul, Korea}

\vspace{1cm}

{\sc 
S.~Heinemeyer$^1$%
\footnote{
email: Sven.Heinemeyer@cern.ch}%
, M.~Mondrag\'on$^2$%
\footnote{
email: myriam@fisica.unam.mx}%
~and G.~Zoupanos$^3$%
\footnote{
email: George.Zoupanos@cern.ch}
}

\vspace*{1cm}

$^1$Instituto de Fisica de Cantabria (CSIC-UC), Santander,  Spain

\vspace*{0.1cm}

$^2$Inst.~de Fisica, Universidad~Nacional Aut\'onoma de M\'exico, 
    M\'exico 01000, D.F., M\'exico

\vspace*{0.1cm}

$^3$Physics Department, National Technical University of Athens,
    Athens, Greece

\end{center}

\vspace*{0.2cm}

\BC {\bf Abstract} \EC
Finite Unified Theories (FUTs) are $N=1$ supersymmetric Grand Unified
Theories that can be made all-loop finite, leading 
to a severe reduction of the free parameters.
We review the investigation of FUTs based on $SU(5)$ in the context of
low-energy phenomenology observables. 
Using the restrictions from the top and bottom quark masses, it is
possible to discriminate between 
different models. Including further low-energy constraints such as
$B$~physics observables, the bound on the lightest Higgs boson mass
and the cold dark matter density, we derive the predictions for the 
supersymmetric particle spectrum and the prospects for discoveries at
the LHC.

\def\thefootnote{\arabic{footnote}}
\setcounter{footnote}{0}

\newpage

%%%%%%%%%%%%%%%%%%%%%%%%%%%%%%%%%%%%%%%%%%%%%%%%%%%%%%%%%%%%%%%%%%%%%%%%%%%%%%%
%%%%%%%%%%%%%%%%%%%%%%%%%%%%%%%%%%%%%%%%%%%%%%%%%%%%%%%%%%%%%%%%%%%%%%%%%%%%%%%

{
\twocolumn

\title{Selecting Finite Unified Theories with Current Data}


\author{S.~Heinemeyer}{
  address={Instituto de Fisica de Cantabria (CSIC-UC), Santander, Spain}
}

\author{M.~Mondrag\'on}{
  address={Inst.~de Fisica, Universidad~Nacional Aut\'onoma de M\'exico, 
  M\'exico 01000, D.F., M\'exico}
}

\author{G.~Zoupanos}{
  address={Physics Department, National Technical University of Athens,
  Athens, Greece 
}
}

\begin{abstract}
Finite Unified Theories (FUTs) are $N=1$ supersymmetric Grand Unified
Theories that can be made all-loop finite, leading 
to a severe reduction of the free parameters.
We review the investigation of FUTs based on $SU(5)$ in the context of
low-energy phenomenology observables. 
Using the restrictions from the top and bottom quark masses, it is
possible to discriminate between 
different models. Including further low-energy constraints such as
$B$~physics observables, the bound on the lightest Higgs boson mass
and the cold dark matter density, we derive the predictions for the 
supersymmetric particle spectrum and the prospects for discoveries at
the LHC.

\end{abstract}

\maketitle

%%%%%%%%%%%%%%%%%%%%%%%%%%%%%%%%%%%%%%%%%%%%%%%%%%%%%%%%%%%%%%%%%%%%%%%%%%%%%%%
%%%%%%%%%%%%%%%%%%%%%%%%%%%%%%%%%%%%%%%%%%%%%%%%%%%%%%%%%%%%%%%%%%%%%%%%%%%%%%%

\section{INTRODUCTION}

Finite Unified Theories are $N=1$ supersymmetric 
GUTs which can be made finite even to all-loop orders,
including the soft supersymmetry (SUSY) breaking sector, see
\citeres{FUTreviews,kkmz1} for details and further references.
The constructed finite unified $N=1$ supersymmetric GUTs
predicted correctly from the dimensionless sector (Gauge-Yukawa
unification), among others, the top quark mass \cite{finite1}.  
Eventually, the predictive power has been extended to the Higgs sector
and the SUSY spectrum. This, in turn, allows to make
predictions for low-energy precision and astrophysical
observables.  

We review an exhaustive search $SU(5)$-based finite SUSY
models, taking into account the restrictions resulting from the
low-energy observables~\cite{fut}%
\footnote{
Analyses of other models based on low-energy precision observables can
be found in \citere{other} and references therein.
}%
.~Finally, the predictions of the
``best'' model (i.e.\ that is still allowed after taking
the phenomenological restrictions into account) for the Higgs and SUSY
searches at the LHC are reviewed~\cite{fut}.

%%%%%%%%%%%%%%%%%%%%%%%%%%%%%%%%%%%%%%%%%%%%%%%%%%%%%%%%%%%%%%%%%%%%%%%%%%%%%%%
%%%%%%%%%%%%%%%%%%%%%%%%%%%%%%%%%%%%%%%%%%%%%%%%%%%%%%%%%%%%%%%%%%%%%%%%%%%%%%%

\section{THE FINITE SU(5) MODELS}

From the classification of finite theories in \citere{Hamidi:1984ft}, 
one can see that there
exist only two candidate possibilities to construct $SU(5)$ GUTs with
three generations. These possibilities require that the theory should
contain as matter fields the chiral super-multiplets 
${\bf 5},~\overline{\bf 5},~{\bf 10}, ~\overline{\bf 5},~{\bf 24}$ with
the multiplicities $(6,9,4,1,0)$ and $(4,7,3,0,1)$, respectively. Only
the second one contains a ${\bf 24}$-plet which can be used to provide
the spontaneous symmetry breaking (SB) of $SU(5)$ down to $SU(3)\times
SU(2)\times U(1)$. Consequently, we concentrate on this class of models.

The particle content of the models we will study consists of the
following super-multiplets: three ($\overline{\bf 5} + \bf{10}$),
needed for each of the three generations of quarks and leptons, four
($\overline{\bf 5} + {\bf 5}$) and one ${\bf 24}$ considered as Higgs
super-multiplets. 
When the gauge group of the finite GUT is broken the theory is no
longer finite, and we will assume that we are left with the MSSM.

In the following we review two versions of the all-order finite
model.  The model of \citere{finite1}, which will be labeled ${\bf
  A}$, and a slight variation of this model (labeled ${\bf B}$), which
can also be obtained from the class of the models suggested in 
\citere{zoup-avdeev1} (with a modification to suppress non-diagonal
anomalous dimensions).
The main difference between model ${\bf A}$ and model ${\bf B}$ is
that two pairs of Higgs quintets and anti-quintets couple to the 
${\bf 24}$ in ${\bf B}$, so that it is not necessary to mix them with
$H_{4}$ and $\overline{H}_{4}$ in order to achieve the triplet-doublet
splitting after the symmetry breaking of $SU(5)$ \cite{kkmz1}.  Thus,
although the particle content is the same, the solutions to restrictions
among the parameters differ, see \citere{fut} for details.
\FUTA\ depends on three free parameters, namely
$m_{\overline{{\bf 5}}}\equiv m_{\overline{{\bf 5}}_3}$, 
$m_{{\bf 10}}\equiv m_{{\bf 10}_3}$
and $M$. 
Here $m_{\bf N}$ denote scalar mass parameters and $M$ the fermionic
mass parameters at the GUT scale.
\FUTB\ is even more restricted and depends only on two free parameters, 
$m_{{\bf 10}}\equiv m_{{\bf 10}_3}$  and $M$ for the dimensionful
sector.

%%%%%%%%%%%%%%%%%%%%%%%%%%%%%%%%%%%%%%%%%%%%%%%%%%%%%%%%%%%%%%%%%%%%%%%%%%%%%%%
%%%%%%%%%%%%%%%%%%%%%%%%%%%%%%%%%%%%%%%%%%%%%%%%%%%%%%%%%%%%%%%%%%%%%%%%%%%%%%%

\section{THE QUARK MASSES}

%%%%%%%%%%%%%%%%%%%%%%%%% F I G U R E %%%%%%%%%%%%%%%%%%%%%%%%%%%%%%%%%%%%%%%%%
\begin{figure}[htb!]
%\vspace{1cm}
\includegraphics[width=0.48\textwidth]{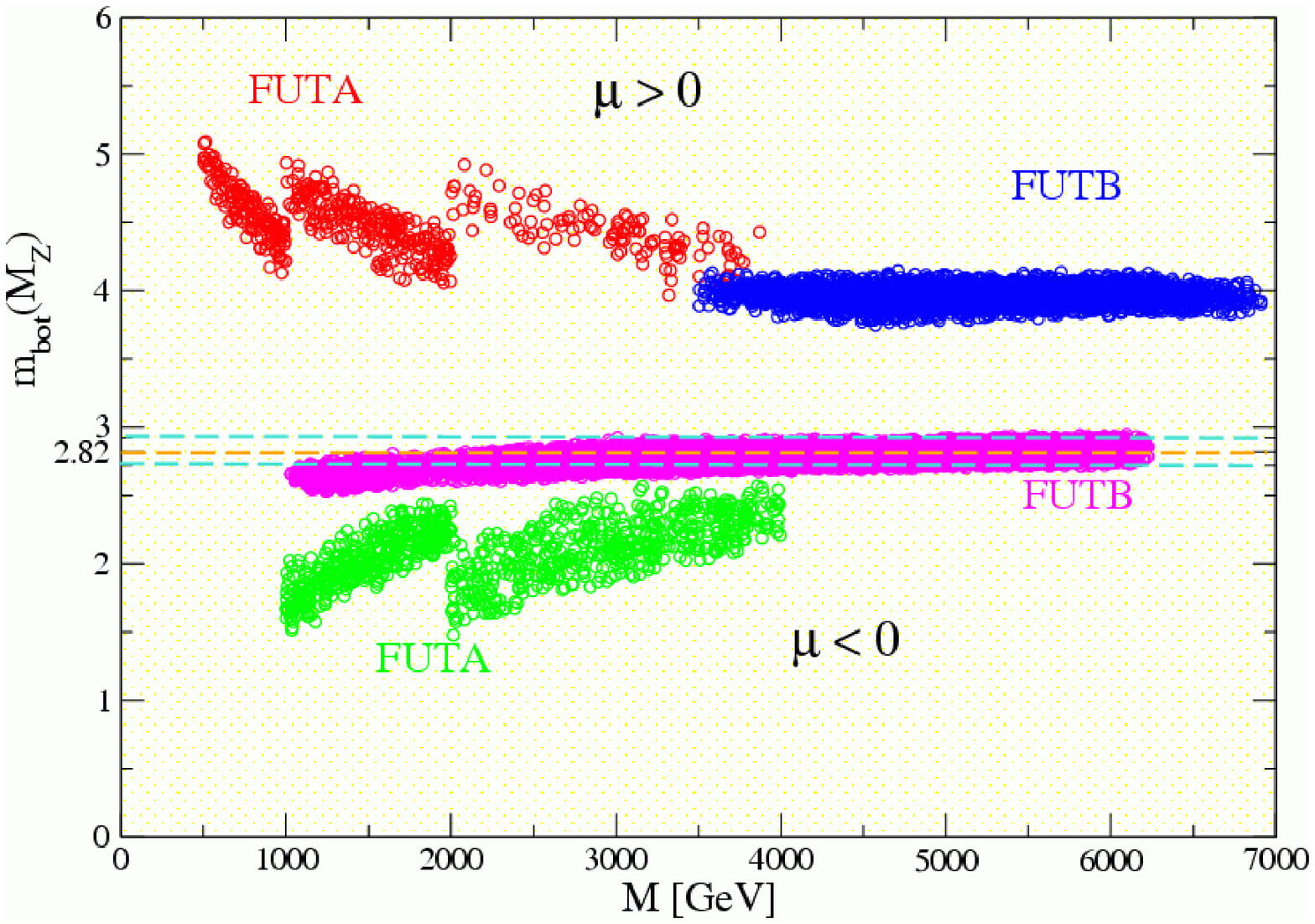}
\includegraphics[width=0.48\textwidth]{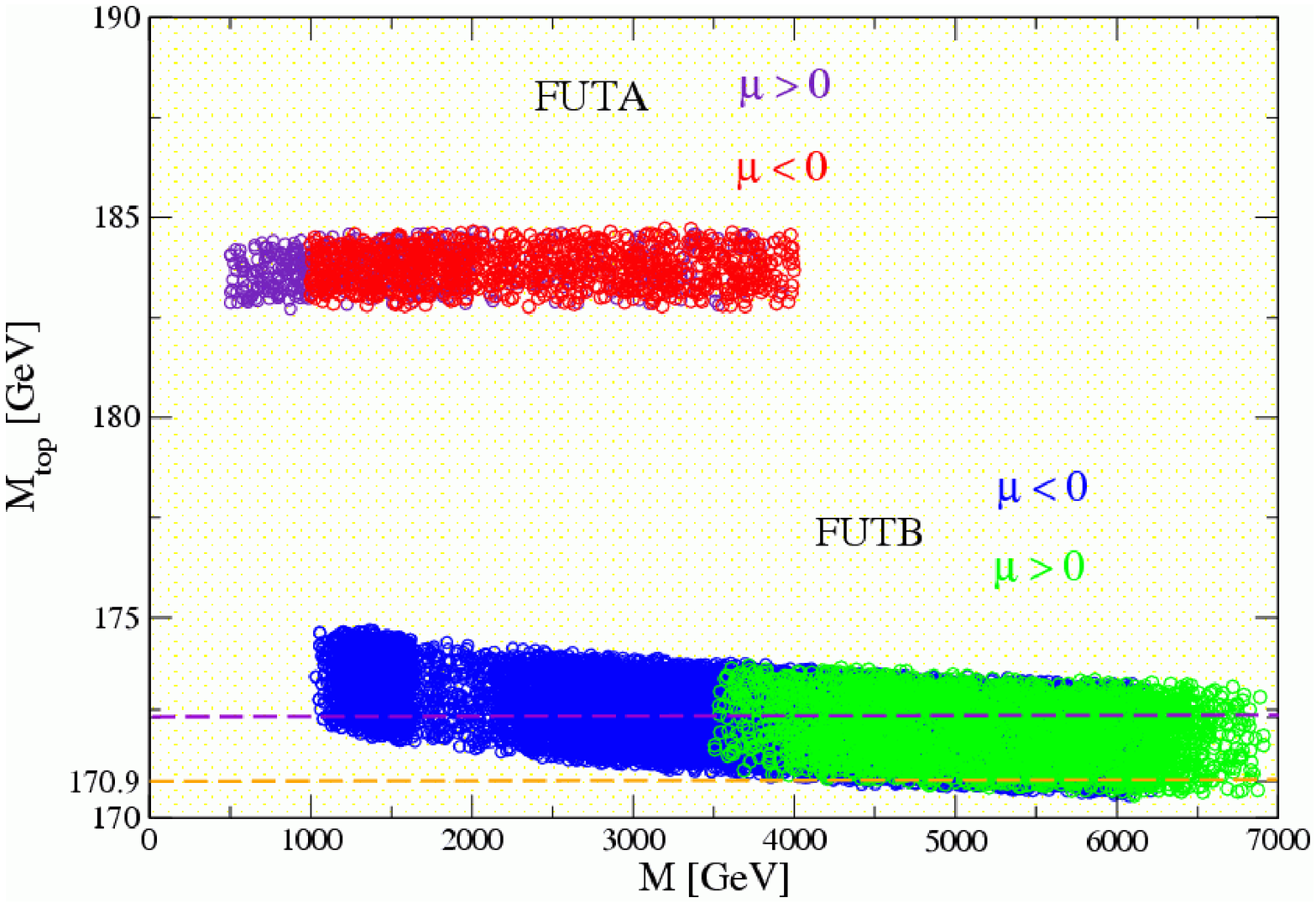}
\caption{$\mbbar(\MZ)$ as function of $M$ (left) and 
$\mt$  as function of $M$ (right) for models $\bf FUTA$ and
$\bf FUTB$, for $\mu<0 $ and $\mu >0$.}
\label{fig:MQvsM}
\end{figure}
%%%%%%%%%%%%%%%%%%%%%%%%% F I G U R E %%%%%%%%%%%%%%%%%%%%%%%%%%%%%%%%%%%%%%%%%

We start with the comparison of the predictions of the two models,
\FUTA\ and \FUTB, for the top- and bottom quark masses with their
experimental values.
For the top-quark mass we used the (older) 
experimental value for the pole mass~\cite{mt1709}
\BE
\label{mtexp}
\mt^{\rm exp} = 170.9 \pm 1.8 \gev~.
\EE
For the bottom-quark mass we use the value at the bottom-quark mass
scale or at $\MZ$~\cite{pdg} 
\BE
\mbbar(\MZ) = 2.82 \pm 0.07 \gev~.
\EE

In \reffi{fig:MQvsM} we present the predictions of the models
concerning the bottom quark mass (left) and the top quark mass (right).
The steps in the values for \FUTA\ 
are due to the fact that fixed values of $M$ were taken, while the
other parameters $m_5$ and $m_{10}$ were varied. However, this selected
sampling of the parameter space is sufficient for us to draw our
conclusions, see below.
We present the predictions for
$\mbbar(M_Z)$, to avoid unnecessary errors coming from the running
from $\MZ$ to the $\mb$ pole mass, which are not related to the
predictions of the present models. The so-called $\db$~effects are taken
into account, see \citere{fut} for more details.
The bounds on the $\mbbar(\MZ)$ and the top quark mass single out \FUTB\
with $\mu < 0$ as the most favorable solution. 
The favored parameter points have $\mu = {\cal O}(-2000 \gev)$ and 
$\tb =$\,\order{50}.

Looking at the anomalous magnetic moment of the
muon~\cite{g-2exp,g-2theo,g-2reviewDS} (see \citere{fut} for a full list
of references) it is obvious that $\mu < 0$ is already
challenged by the present data.
However, a heavy SUSY spectrum as we have here (see below)
with $\mu < 0$ results in a $\amu^{\rm SUSY}$
prediction very close to the SM result. Since the SM is not regarded as
excluded by $(g-2)_\mu$, we continue with our analysis
of \FUTB\ with $\mu < 0$.

%%%%%%%%%%%%%%%%%%%%%%%%%%%%%%%%%%%%%%%%%%%%%%%%%%%%%%%%%%%%%%%%%%%%%%%%%%%%%%%
%%%%%%%%%%%%%%%%%%%%%%%%%%%%%%%%%%%%%%%%%%%%%%%%%%%%%%%%%%%%%%%%%%%%%%%%%%%%%%%

\section{LHC PREDICTIONS}

We now turn to the predictions for the LHC based on model \FUTB\ with
$\mu < 0$. Further experimental bounds applied are the ones of 
$\br(b \to s \ga)$ and $\br(B_s \to \mu^+\mu^-)$ (evaluated with 
{\tt Micromegas}~\cite{micromegas}), the LEP Higgs
bounds~\cite{LEPHiggs} on the lightest Higgs boson mass (evaluated with 
{\tt FeynHiggs}~\cite{feynhiggs,mhiggslong,mhiggsAEC,mhcMSSMlong}) and
the abundance of cold dark matter (CDM) in the early universe (evaluated
with {\tt Micromegas}).

%%%%%%%%%%%%%%%%%%%%%%%%% F I G U R E %%%%%%%%%%%%%%%%%%%%%%%%%%%%%%%%%%%%%%%%%
\begin{figure}[htb!]
%\vspace{1cm}
\includegraphics[width=0.48\textwidth]{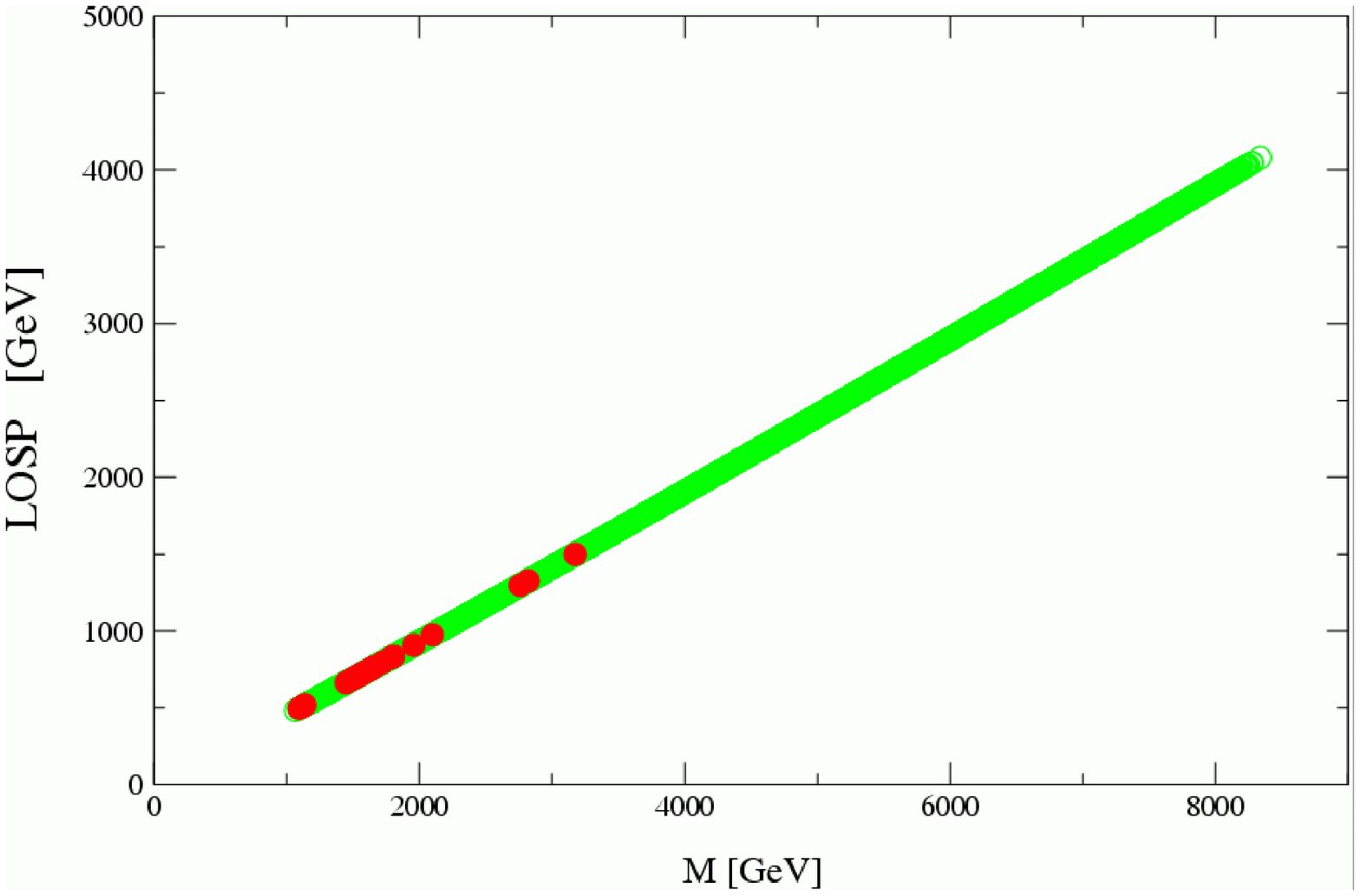}
\includegraphics[width=0.48\textwidth]{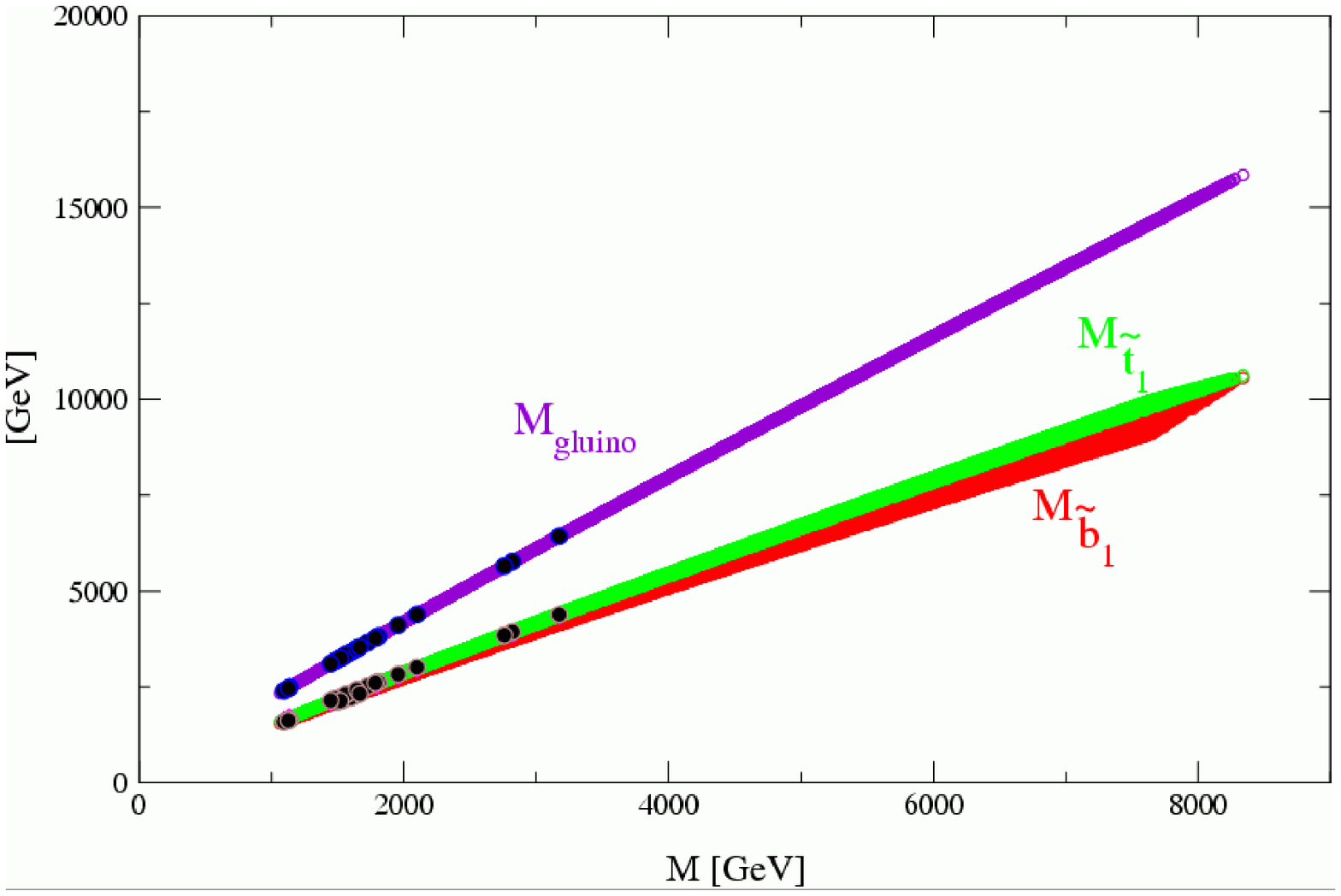}
\caption{
Predictions for SUSY masses of \FUTB\ with $\mu < 0$. The left (right)
plot shows the lightest observable particle (the masses of the lighter
scalar top, scalar bottom and gluino) as a function of~$M$.
The darker shaded (red/left, black/right) dots indicate the points in
agreement with the CDM bound.
}
\label{fig:MvsM}
\end{figure}
%%%%%%%%%%%%%%%%%%%%%%%%% F I G U R E %%%%%%%%%%%%%%%%%%%%%%%%%%%%%%%%%%%%%%%%%

In the left plot of \reffi{fig:MvsM} we show the prediction of \FUTB\
with $\mu < 0$ for the lightest observable particle (LOSP) 
as function of $M$, that comply with the $B$~physics and Higgs constraints.
The darker (red) points fulfill in addition the a loose CDM constraint
(see \citere{fut}). 
The LOSP is either the light scalar $\tau$ or the second lightest
neutralino (which is close in mass with the lightest chargino). 
One can see that the masses are outside the reach of the LHC
and also the ILC. Neglecting the CDM constraint, even higher particle
masses are allowed.

The right plot of \reffi{fig:MvsM} we show the prediction of \FUTB\ with
$\mu < 0$ for the lighter scalar bottom, the lighter scalar top and the
gluino as a function of $M$. The masses show
a nearly linear dependence on $M$. Assuming a discovery reach of 
$\sim 2.5 \tev$ yields a coverage up to $M \lsim 2 \tev$. This
corresponds to the largest part of the CDM favored parameter
space. 
%All these particles are outside the reach of the ILC.
%Disregarding the CDM bounds, see \refse{sec:cdm}, on the other hand,
%results in large parts of the parameter space in which no SUSY
%particle can be observed neither at the LHC nor at the ILC. 

%%%%%%%%%%%%%%%%%%%%%%%%% F I G U R E %%%%%%%%%%%%%%%%%%%%%%%%%%%%%%%%%%%%%%%%%
\begin{figure}[h!]
%\vspace{1cm}
\includegraphics[width=0.48\textwidth]{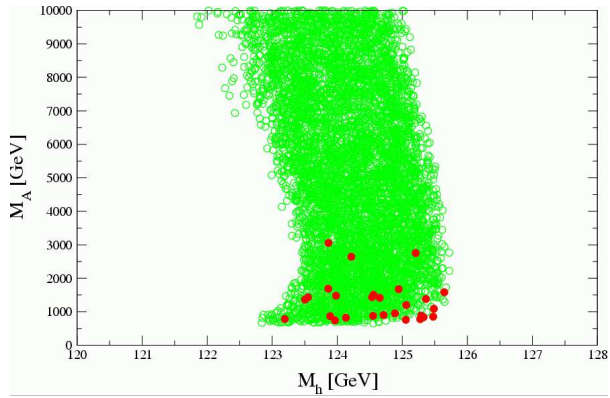}
\caption{
The heavy Higgs boson mass $\MA$ is shown as a function of the lightest
Higgs mass $\Mh$. 
The darker shaded (red) dots indicate the points in
agreement with the CDM bound.
}
\label{fig:MhvsMA}
\end{figure}
%%%%%%%%%%%%%%%%%%%%%%%%% F I G U R E %%%%%%%%%%%%%%%%%%%%%%%%%%%%%%%%%%%%%%%%%

Finally in \reffi{fig:MhvsMA} we show the predictions for the heavy MSSM
Higgs bosons as a function of the lightest Higgs boson mass.
We have truncated the plot at about $\MA = 10 \tev$. The parameter
space allowed by $B$~physics extends up to $\sim 30 \tev$. 
The values that comply with the CDM constraints are in a relatively
light region of $\MA$ with $\MA \lsim 4000 \gev$. 
However, from \ref{fig:MhvsMA} it follows that
the LHC will observe only a light Higgs boson,
whereas the heavy Higgs bosons remain outside the LHC reach, even for
such large values of $|\mu|$ and $\tb$~\cite{cmsreachlargemu}.

%%%%%%%%%%%%%%%%%%%%%%%%%%%%%%%%%%%%%%%%%%%%%%%%%%%%%%%%%%%%%%%%%%%%%%%%%%%%%%%
%%%%%%%%%%%%%%%%%%%%%%%%%%%%%%%%%%%%%%%%%%%%%%%%%%%%%%%%%%%%%%%%%%%%%%%%%%%%%%%

\begin{theacknowledgments}

%This work was supported by the EPEAEK programmes ``Pythagoras'' and
%co-funded by the European Union (75\%) and the Hellenic state (25 \%); also
%supported in part by the mexican grant PAPIIT-UNAM IN115207. 
Partially supported by the NTUA programme for basic research
``K.~Karatheodoris''. 
Work supported in part by the EU's Marie-Curie Research
Training Network under contract MRTN-CT-2006-035505
`Tools and Precision Calculations for Physics Discoveries at Colliders'.
M.M.\ has been supported by the mexican projects PAPIIT IN115207 and
Conacyt  51554-F.

\end{theacknowledgments}

%%%%%%%%%%%%%%%%%%%%%%%%%%%%%%%%%%%%%%%%%%%%%%%%%%%%%%%%%%%%%%%%%%%%%%%%%%%%%%%
%%%%%%%%%%%%%%%%%%%%%%%%%%%%%%%%%%%%%%%%%%%%%%%%%%%%%%%%%%%%%%%%%%%%%%%%%%%%%%%

%\bibliographystyle{aipproc}   % if natbib is available
\bibliographystyle{aipprocl} % if natbib is missing

\begin{thebibliography}{99}

\bibitem{FUTreviews} J.~Kubo, M.~Mondrag\'on and G.~Zoupanos,
                     arXiv:hep-ph/9703289; 
                     %%CITATION = HEP-PH/9703289;%%
                     T.~Kobayashi, J.~Kubo, M.~Mondrag\'on and G.~Zoupanos,
                     {\em AIP Conf.\ Proc.} {\bf 490} (1999) 279.
                     %%CITATION = APCPC,490,279;%%

\bibitem{kkmz1} T.~Kobayashi, J.~Kubo, M.~Mondrag{\'o}n and G.~Zoupanos,
                {\em Nucl. Phys.} {\bf B 511} (1998) 45
                [arXiv:hep-ph/9707425].

\bibitem{finite1} D.~Kapetanakis, M.~Mondrag{\'o}n and G.~Zoupanos, 
                  {\em Zeit. f. Phys.} {\bf C 60} (1993) 181;
                  M.~Mondrag{\'o}n and G.~Zoupanos, 
                  {\em Nucl. Phys.} {\bf C 37} (1995) 98.

\bibitem{fut} S.~Heinemeyer, M.~Mondrag\'on and G.~Zoupanos, 
              {\em JHEP} {\bf 0807} (2008) 135
              [arXiv:0712.3630 [hep-ph]].
              %%CITATION = ARXIV:0712.3630;%%

\bibitem{other} O.~Buchmueller et al.,
                arXiv:0808.4128 [hep-ph];
                %%CITATION = ARXIV:0808.4128;%%
                S.~Heinemeyer, 
                arXiv:0809.2395 [hep-ph].
                %%CITATION = ARXIV:0809.2395;%%

\bibitem{Hamidi:1984ft} S.~Hamidi, J.~Patera and J.~H.~Schwarz,
                        {\em Phys.\ Lett.} {\bf B 141} (1984) 349;
                        %%CITATION = PHLTA,B141,349;%%
                        S.~Rajpoot and J.~Taylor,
                        {\em Phys.\ Lett.} {\bf B 147} (1984) 91;
                        %%CITATION = PHLTA,B147,91;%%
                        X.~Jiang and X.~Zhou,
                        {\em Commun.\ Theor.\ Phys.} {\bf 5} (1986) 179.
                        %%CITATION = CTPMD,5,179;%%

\bibitem{zoup-avdeev1} L.~Avdeev, D.~Kazakov, I.~Kondrashuk,
                       {\em Nucl. Phys.} {\bf B 510} (1998) 289;
                       D.~Kazakov, 
                       {\em Phys. Lett.} {\bf B 449} (1999) 201.

\bibitem{mt1709} Tevatron Electroweak Working Group,
                 hep-ex/0703034.
                 %%CITATION = HEP-EX/0703034;%%

\bibitem{pdg} W.~Yao et al.\  [Particle Data Group Collaboration],
              {\em J.\ Phys.} {\bf G 33} (2006) 1.
              %%CITATION = JPHGB,G33,1;%%

\bibitem{g-2exp} G.~Bennett et al.\ [The Muon g-2 Collaboration],
                 {\em Phys.\ Rev.} {\bf D 73} (2006) 072003
                 [arXiv:hep-ex/0602035].
                 %%CITATION = HEP-EX 0602035;%%

\bibitem{g-2theo} T.~Moroi,
                  {\em Phys. Rev.} {\bf D 53} (1996) 6565
                  [Erratum-ibid.\ {\bf D 56} (1997) 4424]
                  [arXiv:hep-ph/9512396].
                  %%CITATION = HEP-PH 9512396;%%

\bibitem{g-2reviewDS} D.~St\"ockinger,
                      {\em J.\ Phys.} {\bf G 34} (2007) R45
                      [arXiv:hep-ph/0609168].
                      %%CITATION = JPHGB,G34,R45;%%

\bibitem{micromegas} G.~Belanger, F.~Boudjema, A.~Pukhov and A.~Semenov,
                   {\em Comput. Phys. Commun.} {\bf 149} (2002) 103
                   [arXiv:hep-ph/0112278];
                   %%CITATION = HEP-PH 0112278;%%
                   {\em Comput.\ Phys.\ Commun.} {\bf 174} (2006) 577
                   [arXiv:hep-ph/0405253].
                   %%CITATION = HEP-PH 0405253;%%

\bibitem{LEPHiggs} LEP Higgs working group,
                   {\em Phys. Lett.} {\bf B 565} (2003) 61
                   [arXiv:hep-ex/0306033];
                   %%CITATION = HEP-EX 0306033;%%
                   {\em Eur.\ Phys.\ J.} {\bf C 47} (2006) 547
                   [arXiv:hep-ex/0602042].
                   %%CITATION = HEP-EX 0602042;%%

\bibitem{feynhiggs} S.~Heinemeyer, W.~Hollik and G.~Weiglein, 
                    {\em Comp. Phys. Commun.} {\bf 124} 2000 76
                    [arXiv:hep-ph/9812320].
                    %%CITATION = HEP-PH 9812320;%%
                    The code is accessible via
                    {\tt http://www.feynhiggs.de} .

\bibitem{mhiggslong} S.~Heinemeyer, W.~Hollik and G.~Weiglein,
                     {\em Eur. Phys. J.} {\bf C 9} (1999) 343
                     [arXiv:hep-ph/9812472].
                     %%CITATION = HEP-PH 9812472;%%

\bibitem{mhiggsAEC} G.~Degrassi, S.~Heinemeyer, W.~Hollik,
                    P.~Slavich, G.~Weiglein, 
                    {\em Eur. Phys. J.} {\bf C 28} (2003) 133
                    [arXiv:hep-ph/0212020].
                    %%CITATION = HEP-PH 0212020;%%

\bibitem{mhcMSSMlong} M.~Frank et al.,
                      {\em JHEP} {\bf 0702} (2007) 047
                      [arXiv:hep-ph/0611326].
                      %%CITATION = HEP-PH 0611326;%%

\bibitem{cmsreachlargemu} S.~Gennai et al.,
                   {\em Eur.\ Phys.\ J.} {\bf C 52} (2007) 383
                   [arXiv:0704.0619 [hep-ph]];
                   %%CITATION = ARXIV:0704.0619;%%
                   M.~Hashemi et al.,
                   arXiv:0804.1228 [hep-ph];
                   %%CITATION = ARXIV:0804.1228;%%
                   S.~Heinemeyer, A.~Nikitenko and G.~Weiglein,
                   arXiv:0809.2396 [hep-ph];
                   %%CITATION = ARXIV:0809.2396;%%



\end{thebibliography}

%%%%%%%%%%%%%%%%%%%%%%%%%%%%%%%%%%%%%%%%%%%%%%%%%%%%%%%%%%%%%%%%%%%%%%%%%%%%%%%
%%%%%%%%%%%%%%%%%%%%%%%%%%%%%%%%%%%%%%%%%%%%%%%%%%%%%%%%%%%%%%%%%%%%%%%%%%%%%%%

}

\end{document}